\documentclass[aps,prl,preprint,groupedaddress,showpacs]{revtex4-1}
\usepackage{graphicx}
\begin{document}

%\preprint{}

\title{Perfect coupling of light to surface plasmons by coherent absorption}

\author{Heeso Noh}
\author{Yidong Chong}
\author{A. Douglas Stone}
\author{Hui Cao}
\email{hui.cao@yale.edu}
\affiliation{Department of Applied Physics, Yale University, New Haven, CT 06511}



%\thanks{}


%\date{\today}

\begin{abstract}
We show theoretically that coherent light can be completely absorbed
in a two-dimensional or three-dimensional metallic nanostructure by matching the
frequency and field pattern of an incident wave to that of a localized
surface plasmon resonance. This can be regarded as critical coupling
to a nano-plasmonic cavity, or as an extension of the concept of
time-reversed laser to the spaser.  Light scattering is completely
suppressed via impedance matching to the nano-objects, and the energy of incoming wave
is fully transferred to surface plasmon oscillations and evanescent electromagnetic fields. 
Perfect coupling of light to nanostructures has potential applications to
nanoscale probing as well as background-free spectroscopy and
ultrasensitive detection of environmental changes.
\end{abstract}

%73.20.Mf Collective excitations (including excitons, polarons, plasmons and other charge-density excitations)
%42.25.Bs Wave propagation, transmission and absorption
%42.25.Hz interference
%\keywords{}

\maketitle


A fundamental issue in nanophotonics is the efficient delivery of
light into regions with subwavelength dimension, to enhance linear and
nonlinear optical processes on the nanoscale.  This is a formidable
problem because light can normally only be focused down to microscale
regions due to the diffraction limit.  Various schemes have been
developed to couple laser radiation to the nanoscale, e.g.~using
tapered optical fibers or metal tips.  Typically, only a small
fraction of the incident energy can be transferred to the local field,
and the rest is scattered as a stray background.  Moreover, it is
usually impossible to excite a single desired mode of the
nanostructure.  Several recently-proposed schemes include using a
tightly-focused beam to improve the coupling of light with a nanoscale
object \cite{Zumofen2008,Mojarad_OE08}, and enhancing optical
absorption by matching the shape of a nanoparticle to the field
structure of a tightly focused beam \cite{Normatov_OE11}.  While
significant improvements are demonstrated in numerical simulations,
perfect coupling or absorption is still out of the reach.  In this paper,
we propose a method for full delivery of optical energy to individual
resonances of subwavelength structures. 

Our method is based on time-reversed lasing, or coherent perfect
absorption (CPA), a generalization of the concept of critical coupling
that has recently been developed and realized using a simple optical
cavity \cite{Chong_PRL10, Wan_Science11}.  An ordinary cavity made of
dielectric material cannot be much smaller than the wavelength of
light, because resonant feedback cannot be established in a volume of
linear dimension less than a half-wavelength in the medium.  However,
metallic nanostructures support well-known surface plasmon resonances.
Recent work has demonstrated that these resonances can be exploited
in composite structures that include dielectric gain material to
build self-organized oscillations. Such systems, referred to as ``spasers'' (surface plasmon amplification by stimulated emission) \cite{Bergman_PRL03,Zheludev2008,Noginov2009,Oulton2009,Stockman2011,Wuestner2010}, are nanoplasmonic counterpart of lasers (with photons being replaced by surface plasmons and resonant cavities by metallic nanoparticles). 
The cavity modes correspond to localized surface plasmon (LSP) resonances, which can be confined to subwavelength in all three dimensions. 
Surface plasmon oscillations may be coupled to light outside the metallic cavities, producing coherent emission. 
In the time-reversed process, the coherent radiation is switched to an incoming field with phase-conjugated wavefront, and the gain medium replaced by a loss medium.
The time-reversal symmetry demands a complete absorption of the input light, i.e., a perfect conversion of propagating waves from the far-field zone to LSPs and associated evanescent waves in the near-field zone.
The basic theory of coherent perfect absorption applies to spasers, just as to
conventional lasers: if the presence of optical gain can move a
resonant pole of the electromagnetic scattering matrix onto the
real-frequency axis, the time-reverse of the electromagnetic equations
implies that an appropriate amount of loss would produce a zero of the
scattering matrix for real incident electromagnetic waves
\cite{Chong_PRL10, Wan_Science11}.
While the spaser requires a dielectric material with gain, the plasmonic CPA 
does not need a dielectric material with loss, as it can use the intrinsic loss of metal.

The perfect excitation of LSPs in metallic nanostructures is
accompanied by the creation of giant local fields, which are useful
for nanoscale linear and nonlinear optical probing and manipulation.
The fact that incident light can in principle be completely absorbed,
without reradiation, points to possible applications in
background-free spectroscopy and microscopy.  Furthermore, we will see
that the CPA condition can be very sensitive to small changes in the
environment, which points to applications in refractive index sensing
and detection of small concentrations of target molecules. 
The plasmonic CPA (time-reversed spaser) is a general concept for nanoparticles of any shape
or even for clusters, but we will illustrate it here with simple structures such as metallic
nano-cylinders and nano-spheres.

Consider an infinite metallic cylinder of radius $R$ and dielectric
constant $\epsilon_1$, embedded in a dielectric medium $\epsilon_2$.
We denote the cylinder axis by $\hat{z}$.  The incident light is
assumed to propagate in the plane perpendicular to $\hat{z}$, as does
any scattered light.  We take the convention that transverse electric
(TE) polarization denotes an in-plane magnetic field, and transverse
magnetic (TM) polarization denotes an in-plane electric field.  The
two polarizations do not mix upon scattering and can be treated
separately.  TE waves cannot excite LSPs in the metallic cylinder,
because the electric fields are parallel to $\hat{z}$ and do not
generate surface charges.  TM waves, which have magnetic fields
parallel to $\hat{z}$, do couple to LSPs.  Due to the cylindrical
symmetry of the system, the in-plane angular momentum is conserved.
The scattering matrix is therefore diagonal, and its eigenstates
(including zero-scattering eigenstates) have well-defined azimuthal 
number $m$.

Outside the cylinder ($r > R$), the total magnetic field of a TM mode
can be written in cylindrical coordinates as
\begin{equation}
H_z(r, \theta, z) = H^{(2)}_m( n_2 k r) e^{i m \theta} + s
H^{(1)}_m(n_2 k r) e^{i m \theta},
\end{equation}
where $n_2 = \sqrt{\epsilon_2}$, $k = 2 \pi / \lambda$, and $\lambda$
is the vacuum wavelength.  $H^{(1)}_m$ ($H^{(2)}_m$) is the
$m$th-order Hankel function of the first (second) kind, and represents
an outgoing (incoming) wave in the far-field.  $s$ is the scattering
amplitude for this mode.  Inside the metal cylinder ($r < R$), the
magnetic field is $ H_z(r, \theta, z) = a J_m( n_1 k r) e^{i m
  \theta}$, where $J_m$ is the Bessel function of the first kind, $n_1
= \sqrt{\epsilon_1}$, and $a$ is a normalization constant.  By
matching the fields at the metal/dielectric interface ($r = R$), we
get
\begin{equation}
s = {{n_1 J_m(n_1 k R) H^{(2)}_m{'}(n_2 k R) - n_2 J_m'(n_1 k R) H^{(2)}_m( n_2 k R)}
\over{n_2 J_m ' (n_1 k R) H^{(1)}_m( n_2 k R) - n_1 J_m(n_1 k R) H^{(1)}_m{'}( n_2 k R)}},
\label{s}
\end{equation}
where $J'$ ($H'$) is the first-order derivative of the Bessel (Hankel)
function.  For CPA, the scattered wave vanishes ($s = 0$), which
corresponds to the condition
\begin{equation}
n_1 J_m(n_1 k R) H^{(2)}_m{'}( n_2 k R) = n_2 J_m'(n_1 k R) H^{(2)}_m( n_2 k R).
\label{CPAcylinder}
\end{equation}
This is the CPA condition for the whispering-gallery resonances
of a dielectric cylinder.  Now for a metal cylinder with $ Re[\epsilon_1] < 0 $, 
we can obtain solutions with $kR \ll 1$.

\begin{figure}[htbp]
\includegraphics[width=3in]{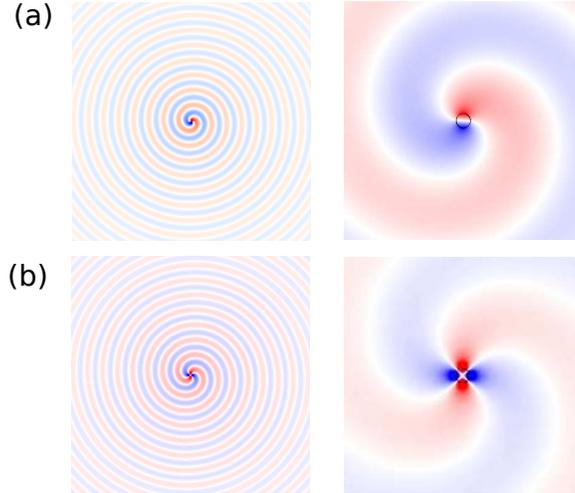}
\caption{Field patterns for CPA with a metallic nano-cylinder.  Left
  column: spatial distribution of magnetic field $H_z$ for perfect
  coupling of a TM-polarized incident wave to a LSP resonance.  For
  (a), $m = 1$ and $\epsilon_1 = -1.14+0.158i$.  For (b), $m=2$ and
  $\epsilon_1 = -1.03+0.00162i$.  In both cases, $kR = 0.3$ and
  $\epsilon_2 = 1$.  Right column: expanded view of the field
  distributions in the vicinity of the cylinder, showing the buildup of
  strong local fields. }
\label{fig1}
\end{figure}

Figure \ref{fig1}(a) plots a solution for $m = 1$, $k R = 0.3$,
$\epsilon_1 = -1.14+0.158i$, and $\epsilon_2 = 1.0$.  The incoming
wave spirals into the metallic cylinder, with zero scattered field.
The radius of the cylinder $R$ is much smaller than $\lambda$.
Figure \ref{fig1}(b) shows another solution for $m = 2$, $k R = 0.3$,
$\epsilon_1 = -1.03+0.00162i$, and $\epsilon_2 = 1.0$.  An expanded
view of the field pattern reveals the buildup of a strong local field
at the cylinder surface. The propagating waves from the far-field are
completely converted to LSP oscillations and evanescent waves in the
near-field.

A similar phenomenon can be realized in a three-dimensional (3D) metal
sphere.  Under spherical symmetry, LSP modes exist only for TM
polarization (the radial component of the magnetic field vanishes).  The CPA condition is obtained following
procedures similar to the above, and can be expressed as
\begin{equation}
\begin{array}{c|c|c}
\epsilon_1 j_l(y_0) \frac{\displaystyle \partial [x h^{(2)}_l(x)]}{\displaystyle \partial x} & \begin{array}{c} \\ _{x = x_0} \end{array} = 
\epsilon_2 h^{(2)}_l(x_0) \frac{\displaystyle \partial [y j_l(y)]}{\displaystyle \partial y} & \begin{array}{c} \\ _{y = y_0} \end{array},
\end{array}
\label{3d cpa}
\end{equation}
where $x = \sqrt{\epsilon_2} \, k \, r $, $y = \sqrt{\epsilon_1} \, k
\, r $, $x_0 = \sqrt{\epsilon_2} \, k \, R $, $y_0 = \sqrt{\epsilon_1}
\, k \, R $, $R$ is the radius of the metal sphere, $\epsilon_1$ and
$\epsilon_2$ are the dielectric constants of the metal sphere and
dielectric host medium respectively.  $j_l$ is the $l$th-order
spherical Bessel function, and $h^{(2)}_l$ the spherical Hankel
function of the second kind.

\begin{figure}[htbp]
\includegraphics[width=3in]{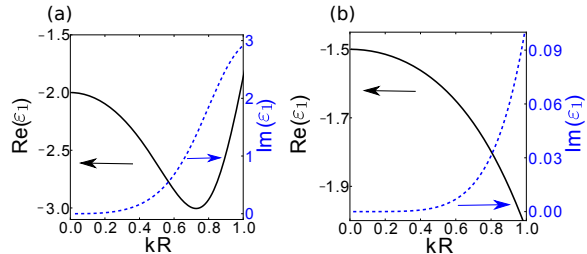}
\caption{CPA solution for a metal sphere in free space. Real part (left vertical axis) and imaginary part (right vertical axis) of the dielectric constant of metal as a function of $kR$ for perfect coupling of impinging light to the LSP modes with $l = 1$ (a) and $l = 2$ (b). CPA is possible for an arbitrary small sphere with finite values of refractive index and absorption coefficient.}
\label{fig2}
\end{figure}

We consider a metallic sphere in free space ($\epsilon_2 = 1.0$).  In
order to generate a CPA resonance at a given frequency, one must tune
both the real part and the imaginary part of $\epsilon_1$ 
\cite{Chong_PRL10}.  Instead of showing specific plasmonic CPA
resonances at given values of $kR$, we show in Fig. \ref{fig2} the
continuous variation of the complex ``CPA dielectric constant'' under
variation of $kR$, for two LSP resonances with $l$ = 1 and 2. 
 As can be seen, plasmonic CPA resonances can be realized for an
arbitrarily small metal sphere.  When $R \rightarrow 0$,
$Re[\epsilon_1] \rightarrow -2.0$ and $Im[\epsilon_1] \rightarrow 0$
for $l = 1$.  This corresponds to the quasi-static limit of the LSP
resonance.  For $l = 2$, $Re[\epsilon_1] \rightarrow -1.5$ and
$Im[\epsilon_1] \rightarrow 0$ as $R \rightarrow 0$.  This solution
again gives a vanishing $Im[\epsilon_1]$ in the quasi-static limit.
As noted above, perfect absorption can be understood as a
generalization of the ``critical coupling'' \cite{Cai_Vahala_PRL00,
  Yariv_PTL02} concept.  For a LSP resonance, critical coupling means
 the rate of radiative loss is equal to that of dissipative loss.
Under time reversal, radiative loss becomes radiative gain, and the
dissipation of energy is fully compensated by the illuminating light.
Since the radiative loss for a quasi-static resonance is extremely
small, the dissipative loss in the nano-sphere would have to be
equally small to achieve perfect absorption.

\begin{figure}[htbp]
\includegraphics[width=2in]{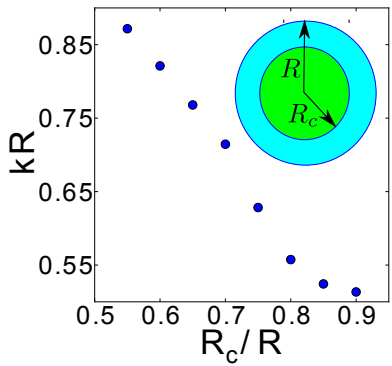}
\caption{CPA for silica core - gold shell structures. Main panel: minimal value of $kR$ as a function of $R_c/R$ for $l = 1$. Inset: a schematic diagram of the core-shell structure. The total structure size $R$ reduces as the metal shell gets thinner.}
\label{fig3}
\end{figure}

Such low dissipative loss is not typically achievable for a solid
metal nanoparticle.  For example, when the dispersive dielectric
constant of gold (Au) \cite{refAu} is inserted into Eq. \ref{3d
  cpa}, one finds that the smallest achievable value of $kR$ is 0.91
at $\lambda = 535$ nm and $l = 2$.  We therefore consider a composite
silica core - gold shell structure, shown schematically in the inset
of Fig. \ref{fig3}.  The core sphere has radius $R_c$ and the shell's
outer radius is $R$.  As the gold shell gets thinner, the fraction of
metal decreases and the dissipative loss of the system is reduced.  As
shown in the main panel of Fig. \ref{fig3}, the minimum achievable
value of $kR$ for CPA decreases as $R_c/R$ increases.  At the same
time, the LSP resonance shifts to longer wavelengths where the metal loss is lower. For example,
when $R_c/R = 0.9$ and $R = 63$ nm, the CPA condition is reached at
$\lambda = 771$ nm.

Although we have only presented calculations for cylindrical and
spherical structures, the time-reversal argument implies that similar
CPA phenomena can occur in more complex structures with LSP
resonances, so long as there is an appropriate amount of dissipative
loss.  It might appear that CPA cannot occur for dark (non-radiative)
plasmonic resonances.  However, a typical dark mode has only vanishing
electric dipole moment; radiation can still occur in higher multipole
moments (e.g.~magnetic dipole and electric quadrupole).  
Hence, it is possible in principle to achieve CPA by coupling to a
dark mode, as long as the radiative coupling of the dark mode is not
completely vanishing; however, the high quality ($Q$) of the
dark mode imposes the requirement of very low dissipation.

Compared to a spaser, a significant difference and also an advantage
of a perfect absorber of the present type is that incident light can
be coupled to any LSP resonance of the structure---not just the
low-loss ones.  Just as in a laser, the spaser usually 
oscillates in the LSP resonances with low loss, which saturate the gain, making it difficult for
higher-loss resonances to reach threshold. In the time-reversed
(absorber) case, the incident wave can be perfectly absorbed by any
LSP resonance so long as its frequency and field pattern match that
particular resonance.  In the regime where absorption is a linear
process, we can simultaneously achieve perfect absorption for several
modes by superimposing the corresponding incident wave patterns.

% The full delivery of optical energy to a nano-object is a result of {\it impedance matching}. 
%Don't like this, impedance matching is a 1d concept and usually refers only to lack of reflection, not full absorption.
%CPA is a 3D multi-channel concept - only for spherically symmetric case does it reduce to one-channel (1D).

\begin{figure}[htbp]
\includegraphics[width=3in]{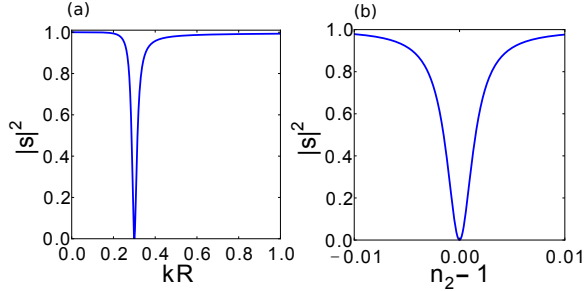}
\caption{Sensitivity of CPA to variations in the incident wave and the
  dielectric environment.  (a) Normalized intensity $|s|^2$ of
  scattered light versus the frequency $k$ of the incident light
  (normalized by $1/R$, where $R$ is the radius of the metallic
  cylinder). CPA is reached at $kR = 0.3$ via resonant excitation of
  the LSP resonance with $m=2$. The spectral width of the dip in
  $|s|^2$ is determined by the dissipative loss of the LSP resonance.
  (b) $|s|^2$ as a function of the change in refractive index $n_2 -1$
  of the dielectric material surrounding the metallic cylinder. CPA
  happens at $n_2 = 1$. A tiny derivation of $n_2$ from one causes a
  dramatic growth of the scattered light intensity. }
\label{fig4}
\end{figure}

The CPA phenomenon is extremely sensitive to variations in the
incident wave or the dielectric environment.  Small changes can
violate the perfect-absorption condition, resulting in a dramatic
increase of the scattered intensity. Figure \ref{fig4}(a) plots the
scattered intensity of the outgoing wave, $|s|^2$, for a metallic
cylinder as a function of the incoming light frequency $k$.  As $k$
deviates from a LSP resonance frequency, the scattered light intensity
increases rapidly from zero.  The spectral width of the CPA resonance
is $\Delta k / k \simeq 0.1$, and it is determined by the dissipative
loss of the LSP mode.  This result is promising for application to
background-free spectroscopy.  A similar effect is found for a tiny
change in refractive index or absorption coefficient of the
surrounding material, as shown in Fig. \ref{fig4}(b) for a variation
in $n_2$.  Hence, the effect may also be useful for ultrasensitive
detection of environmental changes.

In the recent development of metamaterial absorbers, impedance
matching has been used to eliminate reflection of a plane wave from
the front surface, with transmission minimized by the use of multiple
absorbing layers or a mirror at the back surface of the sample
\cite{Landy_PRL08, Avitzour_PRB09, Wang_Soukoulis_PRB09,
  Liu_Giessen_NL10, Polyakov_APL11}.  Although the dimension of each
unit cell is subwavelength, the entire medium is macroscopic. 
CPA, which is a generalization of the concept of impedance matching to arbitrary geometries and dimension,
is applied here to a single subwavelength object in free space.
 By matching the incident field
pattern to the radiation pattern of a LSP resonance, $100\%$ coupling
efficiency is reached, and light scattering in all directions is
eliminated.  However, the input field for perfect-coupling to a nanosphere is an angular
momentum eigenstate, which must converge onto it from all
directions, something not easy to do experimentally.  To
simplify the experimental requirements, it will likely be useful to
strongly break the rotation symmetry, e.g.~by employing ellipsoids or
nano-rods, in order to approach the resonance condition with a more
directional input wave pattern.  Similar directional emission of
lasing from deformed dielectric microcavities has been extensively
studied \cite{NoeckelNature97,JanWiersigPRL08,QinghaiSongPRL10}, which 
could benefit the design of metallic nanoabsorbers.

The CPA mechanism differs from the spatial-temporal localization of
optical energy in a nanoplasmonic system by time-reversal
\cite{Li_Stockman_PRB08}.  In the latter case, a short pulse is
launched from a point in the near field and the radiation in the far
field is recorded.  Time-reversal of the radiated pulse leads to
concentration of input energy at the position of the initial source at
a time corresponding to the end of the excitation pulse.  Afterwards,
the field passes through the focal point and diverges.  The input
energy is not completely absorbed at the focus, unless a time-reversed
source is placed there \cite{deRosny_Fink_PRL02}. In contrast, the CPA
mechanism does not need a coherently-driven source.  Instead it sets
up a perfect ``trap'' for the incident light by matching it to a
resonance of the system.  Thus it works only for coherent light at the
resonant frequency.

In summary, we have demonstrated the possibility of coherent perfect absorption of light by nano-scale metallic objects, equivalent to time-reversing the spaser and perfect coupling
to localized surface plasmons.  The required input field must be monochromatic and
coherent, and the system must have an optimal loss equivalent to the
critical coupling condition.  The required incident waveform depends
on the shape of the nano-scale object; for spherical or cylinderical structures that are considered here,
it is simply a converging spherical or cylindrical TM wave.  Other
structures may require a complex superposition of spherical waves,
the form of which requires further study.

We thank Profs. Mathias Fink, Geoffroy Lerosey, and Rupert F. Oulton for stimulating discussions. 
This work is funded by the NSF Grants DMR-0908437 and ECCS-1068642.

\bibliography{nanoCPA}

%Merlin.mbs v4.21 2009-07-09.
\begin{thebibliography}{10}%
\makeatletter
\providecommand \@ifxundefined [1]{%
 \ifx #1\undefined \expandafter \@firstoftwo
 \else \expandafter \@secondoftwo
\fi
}%
\providecommand \@ifnum [1]{%
 \ifnum #1\expandafter \@firstoftwo
 \else \expandafter \@secondoftwo
\fi
}%
\providecommand \enquote [1]{``#1''}%
\providecommand \bibnamefont  [1]{#1}%
\providecommand \bibfnamefont [1]{#1}%
\providecommand \citenamefont [1]{#1}%
\providecommand\href[0]{\@sanitize\@href}%
\providecommand\@href[1]{\endgroup\@@startlink{#1}\endgroup\@@href}%
\providecommand\@@href[1]{#1\@@endlink}%
\providecommand \@sanitize [0]{\begingroup\catcode`\&12\catcode`\#12\relax}%
\@ifxundefined \pdfoutput {\@firstoftwo}{%
 \@ifnum{\z@=\pdfoutput}{\@firstoftwo}{\@secondoftwo}%
}{%
 \providecommand\@@startlink[1]{\leavevmode\special{html:<a href="#1">}}%
 \providecommand\@@endlink[0]{\special{html:</a>}}%
}{%
 \providecommand\@@startlink[1]{%
  \leavevmode
  \pdfstartlink
   attr{/Border[0 0 1 ]/H/I/C[0 1 1]}%
   user{/Subtype/Link/A<</Type/Action/S/URI/URI(#1)>>}%
  \relax
 }%
 \providecommand\@@endlink[0]{\pdfendlink}%
}%
\providecommand \url  [0]{\begingroup\@sanitize \@url }%
\providecommand \@url [1]{\endgroup\@href {#1}{\urlprefix}}%
\providecommand \urlprefix [0]{URL }%
\providecommand \Eprint[0]{\href }%
\@ifxundefined \urlstyle {%
  \providecommand \doi [1]{doi:\discretionary{}{}{}#1}%
}{%
  \providecommand \doi [0]{doi:\discretionary{}{}{}\begingroup
  \urlstyle{rm}\Url }%
}%
\providecommand \doibase [0]{http://dx.doi.org/}%
\providecommand \Doi[1]{\href{\doibase#1}}%
\providecommand \bibAnnote [3]{%
  \BibitemShut{#1}%
  \begin{quotation}\noindent
    \textsc{Key:}\ #2\\\textsc{Annotation:}\ #3%
  \end{quotation}%
}%
\providecommand \bibAnnoteFile [2]{%
  \IfFileExists{#2}{\bibAnnote {#1} {#2} {\input{#2}}}{}%
}%
\providecommand \typeout [0]{\immediate \write \m@ne }%
\providecommand \selectlanguage [0]{\@gobble}%
\providecommand \bibinfo [0]{\@secondoftwo}%
\providecommand \bibfield [0]{\@secondoftwo}%
\providecommand \translation [1]{[#1]}%
\providecommand \BibitemOpen[0]{}%
\providecommand \bibitemStop [0]{}%
\providecommand \bibitemNoStop [0]{.\EOS\space}%
\providecommand \EOS [0]{\spacefactor3000\relax}%
\providecommand \BibitemShut [1]{\csname bibitem#1\endcsname}%
%</preamble>
\bibitem{Zumofen2008}%
  \BibitemOpen
  \bibfield{author}{%
  \bibinfo {author} {\bibfnamefont{G.}~\bibnamefont{Zumofen}}, \bibinfo
  {author} {\bibfnamefont{N.~M.}\ \bibnamefont{Mojarad}}, \bibinfo {author}
  {\bibfnamefont{V.}~\bibnamefont{Sandoghdar}},\ and\ \bibinfo {author}
  {\bibfnamefont{M.}~\bibnamefont{Agio}},\ }%
  \bibfield{journal}{%
  \bibinfo {journal} {Phys. Rev. Lett.}\ }%
  \textbf{\bibinfo {volume} {101}},\ \bibinfo {pages} {180404} (\bibinfo {year}
  {2008})%
  \bibAnnoteFile{NoStop}{Zumofen2008}%
\bibitem{Mojarad_OE08}%
  \BibitemOpen
  \bibfield{author}{%
  \bibinfo {author} {\bibfnamefont{N.~M.}\ \bibnamefont{Mojarad}}, \bibinfo
  {author} {\bibfnamefont{V.}~\bibnamefont{Sandoghdar}},\ and\ \bibinfo
  {author} {\bibfnamefont{M.}~\bibnamefont{Agio}},\ }%
  \bibfield{journal}{%
  \bibinfo {journal} {J. Opt. Soc. Am. B}\ }%
  \textbf{\bibinfo {volume} {25}},\ \bibinfo {pages} {651} (\bibinfo {year}
  {2008})%
  \bibAnnoteFile{NoStop}{Mojarad_OE08}%
\bibitem{Normatov_OE11}%
  \BibitemOpen
  \bibfield{author}{%
  \bibinfo {author} {\bibfnamefont{A.}~\bibnamefont{Normatov}}, \bibinfo
  {author} {\bibfnamefont{B.}~\bibnamefont{Spektor}}, \bibinfo {author}
  {\bibfnamefont{Y.}~\bibnamefont{Leviatan}},\ and\ \bibinfo {author}
  {\bibfnamefont{J.}~\bibnamefont{Shamir}},\ }%
  \bibfield{journal}{%
  \bibinfo {journal} {Opt. Express}\ }%
  \textbf{\bibinfo {volume} {19}},\ \bibinfo {pages} {8506} (\bibinfo {year}
  {2011})%
  \bibAnnoteFile{NoStop}{Normatov_OE11}%
\bibitem{Chong_PRL10}%
  \BibitemOpen
  \bibfield{author}{%
  \bibinfo {author} {\bibfnamefont{Y.~D.}\ \bibnamefont{Chong}}, \bibinfo
  {author} {\bibfnamefont{L.}~\bibnamefont{Ge}}, \bibinfo {author}
  {\bibfnamefont{H.}~\bibnamefont{Cao}},\ and\ \bibinfo {author}
  {\bibfnamefont{A.~D.}\ \bibnamefont{Stone}},\ }%
  \bibfield{journal}{%
  \bibinfo {journal} {Phys. Rev. Lett.}\ }%
  \textbf{\bibinfo {volume} {105}},\ \bibinfo {pages} {053901} (\bibinfo {year}
  {2010})%
  \bibAnnoteFile{NoStop}{Chong_PRL10}%
\bibitem{Wan_Science11}%
  \BibitemOpen
  \bibfield{author}{%
  \bibinfo {author} {\bibfnamefont{W.}~\bibnamefont{Wan}}, \bibinfo {author}
  {\bibfnamefont{Y.}~\bibnamefont{Chong}}, \bibinfo {author}
  {\bibfnamefont{L.}~\bibnamefont{Ge}}, \bibinfo {author}
  {\bibfnamefont{H.}~\bibnamefont{Noh}}, \bibinfo {author}
  {\bibfnamefont{A.~D.}\ \bibnamefont{Stone}},\ and\ \bibinfo {author}
  {\bibfnamefont{H.}~\bibnamefont{Cao}},\ }%
  \bibfield{journal}{%
  \bibinfo {journal} {Science}\ }%
  \textbf{\bibinfo {volume} {331}},\ \bibinfo {pages} {889} (\bibinfo {year}
  {2011})%
  \bibAnnoteFile{NoStop}{Wan_Science11}%
\bibitem{Bergman_PRL03}%
  \BibitemOpen
  \bibfield{author}{%
  \bibinfo {author} {\bibfnamefont{D.~J.}\ \bibnamefont{Bergman}}\ and\
  \bibinfo {author} {\bibfnamefont{M.~I.}\ \bibnamefont{Stockman}},\ }%
  \bibfield{journal}{%
  \bibinfo {journal} {Phys. Rev. Lett.}\ }%
  \textbf{\bibinfo {volume} {90}},\ \bibinfo {pages} {027402} (\bibinfo {year}
  {2003})%
  \bibAnnoteFile{NoStop}{Bergman_PRL03}%
\bibitem{Zheludev2008}%
  \BibitemOpen
  \bibfield{author}{%
  \bibinfo {author} {\bibfnamefont{N.~I.}\ \bibnamefont{Zheludev}}, \bibinfo
  {author} {\bibfnamefont{L.~S.}\ \bibnamefont{Prosvirnin}}, \bibinfo {author}
  {\bibfnamefont{N.}~\bibnamefont{Papasimakis}},\ and\ \bibinfo {author}
  {\bibfnamefont{A.~V.}\ \bibnamefont{Fedotov}},\ }%
  \bibfield{journal}{%
  \bibinfo {journal} {Nature Photon.}\ }%
  \textbf{\bibinfo {volume} {2}},\ \bibinfo {pages} {351} (\bibinfo {year}
  {2008})%
  \bibAnnoteFile{NoStop}{Zheludev2008}%
\bibitem{Noginov2009}%
  \BibitemOpen
  \bibfield{author}{%
  \bibinfo {author} {\bibfnamefont{M.~A.}\ \bibnamefont{Noginov}}, \bibinfo
  {author} {\bibfnamefont{G.}~\bibnamefont{Zhu}}, \bibinfo {author}
  {\bibfnamefont{A.~M.}\ \bibnamefont{Belgrave}}, \bibinfo {author}
  {\bibfnamefont{R.}~\bibnamefont{Bakker}}, \bibinfo {author}
  {\bibfnamefont{V.~M.}\ \bibnamefont{Shalaev}}, \bibinfo {author}
  {\bibfnamefont{E.~E.}\ \bibnamefont{Narimanov}}, \bibinfo {author}
  {\bibfnamefont{S.}~\bibnamefont{Stout}}, \bibinfo {author}
  {\bibfnamefont{E.}~\bibnamefont{Herz}}, \bibinfo {author}
  {\bibfnamefont{T.}~\bibnamefont{Suteewong}},\ and\ \bibinfo {author}
  {\bibfnamefont{U.}~\bibnamefont{Wiesner}},\ }%
  \bibfield{journal}{%
  \bibinfo {journal} {Nature}\ }%
  \textbf{\bibinfo {volume} {460}},\ \bibinfo {pages} {1110} (\bibinfo {year}
  {2009})%
  \bibAnnoteFile{NoStop}{Noginov2009}%
\bibitem{Oulton2009}%
  \BibitemOpen
  \bibfield{author}{%
  \bibinfo {author} {\bibfnamefont{R.~F.}\ \bibnamefont{Oulton}}, \bibinfo
  {author} {\bibfnamefont{V.~J.}\ \bibnamefont{Sorger}}, \bibinfo {author}
  {\bibfnamefont{T.}~\bibnamefont{Zentgraf}}, \bibinfo {author}
  {\bibfnamefont{R.-M.}\ \bibnamefont{Ma}}, \bibinfo {author}
  {\bibfnamefont{C.}~\bibnamefont{Gladden}}, \bibinfo {author}
  {\bibfnamefont{L.}~\bibnamefont{Dai}}, \bibinfo {author}
  {\bibfnamefont{G.}~\bibnamefont{Bartal}},\ and\ \bibinfo {author}
  {\bibfnamefont{X.}~\bibnamefont{Zhang}},\ }%
  \bibfield{journal}{%
  \bibinfo {journal} {Nature}\ }%
  \textbf{\bibinfo {volume} {461}},\ \bibinfo {pages} {629} (\bibinfo {year}
  {2009})%
  \bibAnnoteFile{NoStop}{Oulton2009}%
\bibitem{Stockman2011}%
  \BibitemOpen
  \bibfield{author}{%
  \bibinfo {author} {\bibfnamefont{M.~I.}\ \bibnamefont{Stockman}},\ }%
  \bibfield{journal}{%
  \bibinfo {journal} {Phys. Rev. Lett.}\ }%
  \textbf{\bibinfo {volume} {106}},\ \bibinfo {pages} {156802} (\bibinfo {year}
  {2011})%
  \bibAnnoteFile{NoStop}{Stockman2011}%
\bibitem{Wuestner2010}%
  \BibitemOpen
  \bibfield{author}{%
  \bibinfo {author} {\bibfnamefont{S.}~\bibnamefont{Wuestner}}, \bibinfo
  {author} {\bibfnamefont{A.}~\bibnamefont{Pusch}}, \bibinfo {author}
  {\bibfnamefont{K.~L.}\ \bibnamefont{Tsakmakidis}}, \bibinfo {author}
  {\bibfnamefont{J.~M.}\ \bibnamefont{Hamm}},\ and\ \bibinfo {author}
  {\bibfnamefont{O.}~\bibnamefont{Hess}},\ }%
  \bibfield{journal}{%
  \bibinfo {journal} {Phys. Rev. Lett.}\ }%
  \textbf{\bibinfo {volume} {105}},\ \bibinfo {pages} {127401} (\bibinfo {year}
  {2010})%
  \bibAnnoteFile{NoStop}{Wuestner2010}%
\bibitem{Cai_Vahala_PRL00}%
  \BibitemOpen
  \bibfield{author}{%
  \bibinfo {author} {\bibfnamefont{M.}~\bibnamefont{Cai}}, \bibinfo {author}
  {\bibfnamefont{O.}~\bibnamefont{Painter}},\ and\ \bibinfo {author}
  {\bibfnamefont{K.~J.}\ \bibnamefont{Vahala}},\ }%
  \bibfield{journal}{%
  \bibinfo {journal} {Phys. Rev. Lett.}\ }%
  \textbf{\bibinfo {volume} {85}},\ \bibinfo {pages} {74} (\bibinfo {year}
  {2000})%
  \bibAnnoteFile{NoStop}{Cai_Vahala_PRL00}%
\bibitem{Yariv_PTL02}%
  \BibitemOpen
  \bibfield{author}{%
  \bibinfo {author} {\bibfnamefont{A.}~\bibnamefont{Yariv}},\ }%
  \bibfield{journal}{%
  \bibinfo {journal} {IEEE Photon. Technol. Lett.}\ }%
  \textbf{\bibinfo {volume} {14}},\ \bibinfo {pages} {483 } (\bibinfo {year}
  {2002})%
  \bibAnnoteFile{NoStop}{Yariv_PTL02}%
\bibitem{refAu}%
  \BibitemOpen
  \bibfield{author}{%
  \bibinfo {author} {\bibfnamefont{P.~B.}\ \bibnamefont{Johnson}}\ and\
  \bibinfo {author} {\bibfnamefont{R.~W.}\ \bibnamefont{Christy}},\ }%
  \bibfield{journal}{%
  \bibinfo {journal} {Phys. Rev. B}\ }%
  \textbf{\bibinfo {volume} {6}},\ \bibinfo {pages} {4370} (\bibinfo {year}
  {1972})%
  \bibAnnoteFile{NoStop}{refAu}%
\bibitem{Landy_PRL08}%
  \BibitemOpen
  \bibfield{author}{%
  \bibinfo {author} {\bibfnamefont{N.~I.}\ \bibnamefont{Landy}}, \bibinfo
  {author} {\bibfnamefont{S.}~\bibnamefont{Sajuyigbe}}, \bibinfo {author}
  {\bibfnamefont{J.~J.}\ \bibnamefont{Mock}}, \bibinfo {author}
  {\bibfnamefont{D.~R.}\ \bibnamefont{Smith}},\ and\ \bibinfo {author}
  {\bibfnamefont{W.~J.}\ \bibnamefont{Padilla}},\ }%
  \bibfield{journal}{%
  \bibinfo {journal} {Phys. Rev. Lett.}\ }%
  \textbf{\bibinfo {volume} {100}},\ \bibinfo {pages} {207402} (\bibinfo {year}
  {2008})%
  \bibAnnoteFile{NoStop}{Landy_PRL08}%
\bibitem{Avitzour_PRB09}%
  \BibitemOpen
  \bibfield{author}{%
  \bibinfo {author} {\bibfnamefont{Y.}~\bibnamefont{Avitzour}}, \bibinfo
  {author} {\bibfnamefont{Y.~A.}\ \bibnamefont{Urzhumov}},\ and\ \bibinfo
  {author} {\bibfnamefont{G.}~\bibnamefont{Shvets}},\ }%
  \bibfield{journal}{%
  \bibinfo {journal} {Phys. Rev. B}\ }%
  \textbf{\bibinfo {volume} {79}},\ \bibinfo {pages} {045131} (\bibinfo {year}
  {2009})%
  \bibAnnoteFile{NoStop}{Avitzour_PRB09}%
\bibitem{Wang_Soukoulis_PRB09}%
  \BibitemOpen
  \bibfield{author}{%
  \bibinfo {author} {\bibfnamefont{B.}~\bibnamefont{Wang}}, \bibinfo {author}
  {\bibfnamefont{T.}~\bibnamefont{Koschny}},\ and\ \bibinfo {author}
  {\bibfnamefont{C.~M.}\ \bibnamefont{Soukoulis}},\ }%
  \bibfield{journal}{%
  \bibinfo {journal} {Phys. Rev. B}\ }%
  \textbf{\bibinfo {volume} {80}},\ \bibinfo {pages} {033108} (\bibinfo {year}
  {2009})%
  \bibAnnoteFile{NoStop}{Wang_Soukoulis_PRB09}%
\bibitem{Liu_Giessen_NL10}%
  \BibitemOpen
  \bibfield{author}{%
  \bibinfo {author} {\bibfnamefont{N.}~\bibnamefont{Liu}}, \bibinfo {author}
  {\bibfnamefont{M.}~\bibnamefont{Mesch}}, \bibinfo {author}
  {\bibfnamefont{T.}~\bibnamefont{Weiss}}, \bibinfo {author}
  {\bibfnamefont{M.}~\bibnamefont{Hentschel}},\ and\ \bibinfo {author}
  {\bibfnamefont{H.}~\bibnamefont{Giessen}},\ }%
  \bibfield{journal}{%
  \bibinfo {journal} {Nano Lett.}\ }%
  \textbf{\bibinfo {volume} {10}},\ \bibinfo {pages} {2342} (\bibinfo {year}
  {2010})%
  \bibAnnoteFile{NoStop}{Liu_Giessen_NL10}%
\bibitem{Polyakov_APL11}%
  \BibitemOpen
  \bibfield{author}{%
  \bibinfo {author} {\bibfnamefont{A.}~\bibnamefont{Polyakov}}, \bibinfo
  {author} {\bibfnamefont{S.}~\bibnamefont{Cabrini}}, \bibinfo {author}
  {\bibfnamefont{S.}~\bibnamefont{Dhuey}}, \bibinfo {author}
  {\bibfnamefont{B.}~\bibnamefont{Harteneck}}, \bibinfo {author}
  {\bibfnamefont{P.~J.}\ \bibnamefont{Schuck}},\ and\ \bibinfo {author}
  {\bibfnamefont{H.~A.}\ \bibnamefont{Padmore}},\ }%
  \bibfield{journal}{%
  \bibinfo {journal} {Appl. Phys. Lett.}\ }%
  \textbf{\bibinfo {volume} {98}},\ \bibinfo {pages} {203104} (\bibinfo {year}
  {2011})%
  \bibAnnoteFile{NoStop}{Polyakov_APL11}%
\bibitem{NoeckelNature97}%
  \BibitemOpen
  \bibfield{author}{%
  \bibinfo {author} {\bibfnamefont{J.~U.}\ \bibnamefont{Nockel}}\ and\ \bibinfo
  {author} {\bibfnamefont{A.~D.}\ \bibnamefont{Stone}},\ }%
  \bibfield{journal}{%
  \bibinfo {journal} {Nature}\ }%
  \textbf{\bibinfo {volume} {385}},\ \bibinfo {pages} {45} (\bibinfo {year}
  {1997})%
  \bibAnnoteFile{NoStop}{NoeckelNature97}%
\bibitem{JanWiersigPRL08}%
  \BibitemOpen
  \bibfield{author}{%
  \bibinfo {author} {\bibfnamefont{J.}~\bibnamefont{Wiersig}}\ and\ \bibinfo
  {author} {\bibfnamefont{M.}~\bibnamefont{Hentschel}},\ }%
  \bibfield{journal}{%
  \bibinfo {journal} {Phys. Rev. Lett.}\ }%
  \textbf{\bibinfo {volume} {100}},\ \bibinfo {pages} {033901} (\bibinfo {year}
  {2008})%
  \bibAnnoteFile{NoStop}{JanWiersigPRL08}%
\bibitem{QinghaiSongPRL10}%
  \BibitemOpen
  \bibfield{author}{%
  \bibinfo {author} {\bibfnamefont{Q.~H.}\ \bibnamefont{Song}}\ and\ \bibinfo
  {author} {\bibfnamefont{H.}~\bibnamefont{Cao}},\ }%
  \bibfield{journal}{%
  \bibinfo {journal} {Phys. Rev. Lett.}\ }%
  \textbf{\bibinfo {volume} {105}},\ \bibinfo {pages} {053902} (\bibinfo {year}
  {2010})%
  \bibAnnoteFile{NoStop}{QinghaiSongPRL10}%
\bibitem{Li_Stockman_PRB08}%
  \BibitemOpen
  \bibfield{author}{%
  \bibinfo {author} {\bibfnamefont{X.}~\bibnamefont{Li}}\ and\ \bibinfo
  {author} {\bibfnamefont{M.~I.}\ \bibnamefont{Stockman}},\ }%
  \bibfield{journal}{%
  \bibinfo {journal} {Phys. Rev. B}\ }%
  \textbf{\bibinfo {volume} {77}},\ \bibinfo {pages} {195109} (\bibinfo {year}
  {2008})%
  \bibAnnoteFile{NoStop}{Li_Stockman_PRB08}%
\bibitem{deRosny_Fink_PRL02}%
  \BibitemOpen
  \bibfield{author}{%
  \bibinfo {author} {\bibfnamefont{J.}~\bibnamefont{de~Rosny}}\ and\ \bibinfo
  {author} {\bibfnamefont{M.}~\bibnamefont{Fink}},\ }%
  \bibfield{journal}{%
  \bibinfo {journal} {Phys. Rev. Lett.}\ }%
  \textbf{\bibinfo {volume} {89}},\ \bibinfo {pages} {124301} (\bibinfo {year}
  {2002})%
  \bibAnnoteFile{NoStop}{deRosny_Fink_PRL02}%
\end{thebibliography}%

\end{document}